\documentclass{iopart}

\usepackage{graphicx}
\usepackage{amssymb}
\usepackage{color}
\usepackage{ dsfont }

\begin{document}

\title{Topological Qubits with Majorana Fermions in Trapped Ions}

\author{A Mezzacapo$^{1}$, J Casanova$^{1}$, L Lamata$^{1}$, and E Solano$^{1,2}$}

\address{$^1$ Departamento de Qu\'{\i}mica F\'{\i}sica, Universidad del Pa\'{\i}s Vasco UPV/EHU, Apartado 644, 48080 Bilbao, Spain}

\address{$^2$ IKERBASQUE, Basque Foundation for Science, Alameda Urquijo 36, 48011 Bilbao,
Spain}
 \ead{ant.mezzacapo@gmail.com}
\begin{abstract}
We propose a method of encoding a topologically-protected qubit using Majorana fermions in a trapped-ion chain.  This qubit is protected against major sources of decoherence, while local operations and measurements can be realized. Furthermore, we show that an efficient quantum interface and memory for arbitrary multiqubit photonic states can be built, encoding them into a set of entangled Majorana-fermion qubits inside cavities.    
\end{abstract}

\maketitle

\section{Introduction}
The search for a decoherence free qubit is a major challenge in the
field of quantum computation~\cite{Nielsen00}. Topologically protected systems
offer promising properties for the building of a fault-tolerant quantum
memory~\cite{Nayak08}. However, the realization of topological quantum memories up to now represents
a challenging open problem.

The discrete quantum wire model by A. Kitaev is one of the simplest systems supporting topological phases~\cite{KitaevWire01}. In this model, the signature for the topological nontrivial phase is the presence of unpaired Majorana fermions (MFs). Among their properties, these fermions coincide with their own antiparticles and support non-Abelian statistics~\cite{Alicea11}. Since its
appearance, the model has attracted much interest, with diverse proposals for physical implementations,
including superconducting heterostructures~\cite{Lutchin,Oreg10,Hassler12}
and optical lattices~\cite{Jiang11,Diehl11}. Signatures of these particles have been recently measured~\cite{Mourik12}. However, a clear way to use the fault-tolerant properties of Majorana modes has not been experimentally achieved. In general, building a qubit made out of MFs requires exceptional system control.

 Trapped ions are highly controllable quantum systems~\cite{Leibfried03}. They can be cooled down to form crystals, easily initialized by optical pumping, manipulated with lasers, and efficiently measured. They offer one of the most reliable and scalable implementations for a quantum simulator~\cite{Feynman82,Lloyd96}. Some examples are spin systems~\cite{Porras04,Friedenauer08,Kim10,Schmidt-Kaler,Barreiro11,Lanyon11}, relativistic quantum mechanics~\cite{Lamata07,Gerritsma1,Casanova1,Gerritsma2,Casanova11,Lamata11}, quantum field theories~\cite{CasanovaQFT,PreskillQFT}, and fermionic systems~\cite{CasanovaFermions}. A proposal for realizing another topological system, i.e., Kitaev honeycomb model, was put forward in Ref.~\cite{Schmied11}. However, the complexity of the honeycomb model, requiring three different kinds of interactions, $XX$, $YY$, and $ZZ$, is much higher than the one of the wire model, that only requires an $XX$ interaction. Another proposal for topological systems in trapped ions~\cite{Milman}, not involving Majorana fermions, makes use of a highly nonlocal Hamiltonian.

We propose the implementation of Kitaev~\cite{KitaevWire01} wire model in a linear chain of trapped ions. By a mapping the Kitaev Hamiltonian onto a spin model, we show that this system can be realized in a trapped-ion chain with current technology, and a MF qubit can be encoded. This qubit will be topologically protected against major sources of decoherence for longer times, constituting an efficient quantum memory.  The proposed implementation, already valid for $3$ ions, allows for the straightforward realization of local operations on the MF qubit and for an efficient readout of its state. We also show that a quantum interface between highly entangled incoming photonic states and MF qubits can be implemented by grouping many of these basic units. To this end, we suggest the use of an array of ion chains inside a set of cavities as a possible experimental realization.
\section{Trapped-ion implementation}
We begin by considering the Kitaev Hamiltonian~\cite{KitaevWire01} made up of $N$ spinless fermionic
sites ($\hbar=1$),
\begin{eqnarray}
\nonumber H  =\sum_{j=1}^{N-1}[-w(a_{j}^{\dagger}a_{j+1}+a_{j+1}^{\dagger}a_{j})+\Delta a_{j}a_{j+1}+\Delta^{*}a_{j+1}^{\dagger}a_{j}^{\dagger}]-\\ \mu\sum_{j=1}^N\left(a_{j}^{\dagger}a_{j}-\frac{1}{2}\right),\label{KitaevHfermion}
\end{eqnarray}
where the operators $a_i(a_i^{\dagger})$ are the annihilation (creation) operators of spinless fermions on the site $i$, satisfying $ \{a_i,a_j^{\dagger} \}=\delta_{ij} $, $w$ is the hopping energy, $\mu$ is a chemical potential and $\Delta$ is the order parameter of the pairing term,
which mimics a $p$-wave spinless superconductivity. Given that $\Delta=|\Delta|e^{i\phi}$,
one can define a set of $2N$ Majorana fermions $c_{2j-1}=e^{i\frac{\phi}{2}}a_{j}+e^{-i\frac{\phi}{2}}a_{j}^{\dagger}$,
$c_{2j}=-ie^{i\frac{\phi}{2}}a_{j}+ie^{-i\frac{\phi}{2}}a_{j}^{\dagger}$,
and the Hamiltonian becomes
\begin{eqnarray}
H_M & = & \frac{i}{2}\{{-\sum_{j=1}^N}\mu c_{2j-1}c_{2j}+\sum_{j=1}^{N-1}[(w+|\Delta|)c_{2j}c_{2j+1}\nonumber\\&&+(-w+|\Delta|)c_{2j-1}c_{2j+2}]\}.
\end{eqnarray}
Under the parameter regime $\mu=0$, $|\Delta|=w>0$ the Majorana
fermions $c_{1}$ and $c_{2N}$ disappear, and $H_M$ becomes diagonal in the basis $\tilde{a}_{j}=\frac{1}{2}(c_{2j}+ic_{2j+1})$, $j=1,...,N-1$. One can
pair the two outer Majorana fermions into an additional complex fermion $a_{M}^{\dagger}=(c_{1}+i c_{2N})/2$.
The ground state is twofold
degenerate, and is spanned by the states $|\Psi_{0}\rangle$ and
$|\Psi_{1}\rangle$ defined by $ $$\tilde{a}_{i}|\Psi_{0,1}\rangle=0,\:\forall i=1,2,...,N-1$
and $a_{M}|\Psi_{0}\rangle=0$, $|\Psi_{1}\rangle=a_{M}^{\dagger}|\Psi_{0}\rangle$.  This ground state is separated by a gap
$2w$ from the higher-energy excitations of the system.
It was proved~\cite{KitaevWire01} that these states survive in the regime $|\mu|\ll2w$. Under this regime one has a nontrivial value for the topological invariant $\mathds{Z}_2$, that labels the two different topological phases of the ground state.

In order to implement the Kitaev model in an ion string, we take into account that Hamiltonian in Eq. (\ref{KitaevHfermion}), for the parameter regime  $\phi=0$, which does not reduce the generality of the model,  and $|\Delta|=w>0$, can be mapped onto the transverse field Ising model by using a Jordan-Wigner transformation~\cite{JordanWigner,KitaevLaumann}, 
\begin{equation}
H_s=-J\sum_{j=1}^{N-1}\sigma_j^x\sigma_{j+1}^x-h_z\sum_{j=1}^N\sigma_j^z,\label{KitaevSpinH}
\end{equation} 
where $J=w$ is the exchange coupling and $h_z=-\mu/2$ is a transverse magnetic field. The mapping of the MFs onto the spins is $c_{2j}=\prod_{k=1}^{j-1}\sigma_k^z\sigma_j^y$, $c_{2j-1}=\prod_{k=1}^{j-1}\sigma_k^z\sigma_j^x$. The definitions of $a_M,$ $\tilde{a}_i,$ and $|\Psi_{0,1}\rangle$ change accordingly. The two ground states $|\Psi_{0,1}\rangle$ of $H_s$ in Eq. (\ref{KitaevSpinH})  consequently span a subspace that is protected from higher-level excitations by the energy gap $2w=2J$.

Hamiltonian (\ref{KitaevSpinH}) can be implemented with standard trapped-ion technology~\cite{Leibfried03}. We consider a 1D string of $N$ two-level ions coupled through the joint motional modes by means of external lasers. One possibility is to apply Raman lasers for the spin-spin interaction and for the magnetic field at the same time~\cite{Porras04,Friedenauer08,Kim09}. By locally addressing each ion with different Raman beatnotes, in a multiple M\o lmer-S\o rensen configuration, one can achieve the $\sum_{j=1}^{N-1}\sigma_j^x\sigma_{j+1}^x$ interaction with open boundary conditions in Eq. (\ref{KitaevSpinH}), while a single laser can implement the  $ \sum_j\sigma_j^z$ part~\cite{Kim09}. 

\begin{figure}[t] 
\includegraphics[width=0.9\linewidth]{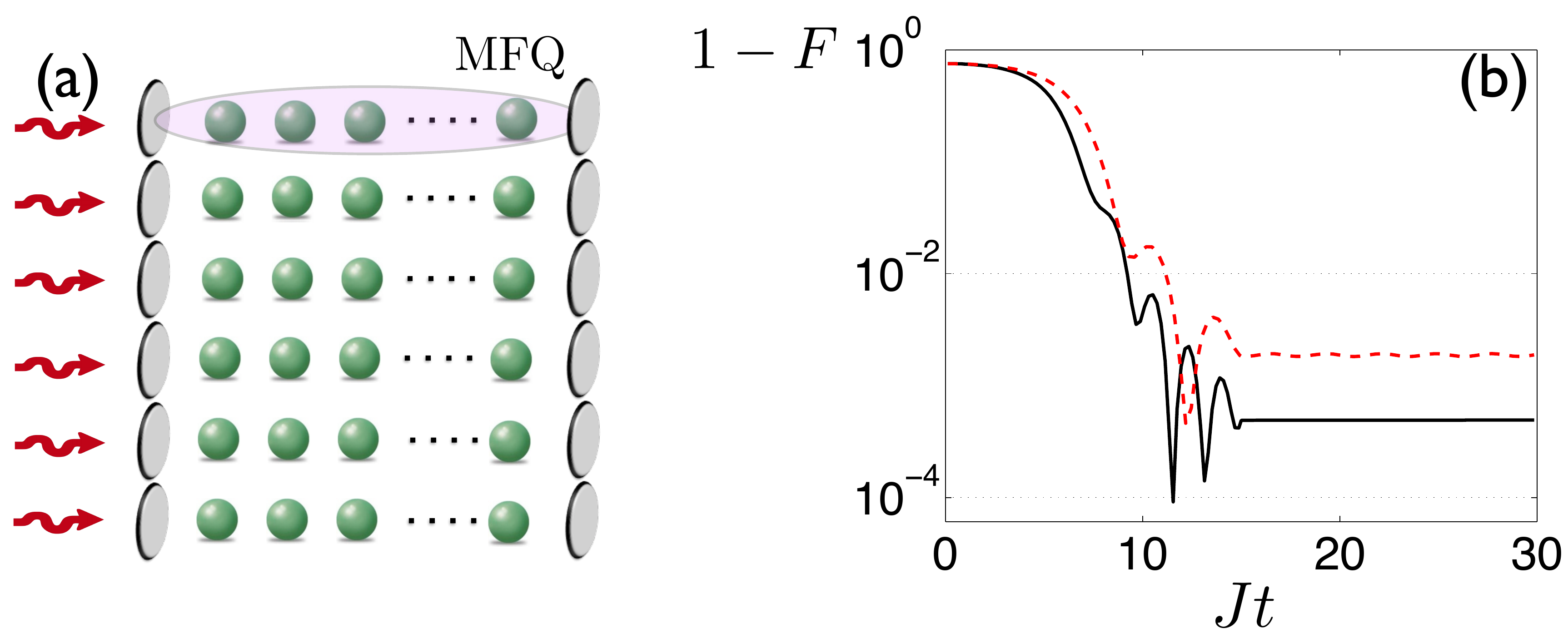}\centering
\caption{(a) Array of 1D ion crystals inside $K$ optical cavities for the quantum interface between a multiqubit entangled photonic state and a MF qubit multipartite state. The shaded region denotes a single MF qubit. (b) Fidelity loss $1-F=1-|\langle \Psi_0|\psi(t)\rangle|^2$ for the evolved state $|\psi(t)\rangle $ from the initial state $|\Psi_0^{\rm init}\rangle$ using the adiabatic transfer protocol with $H_s$ in Eq. (\ref{KitaevSpinH}), for $N=3$. The diabatic error for the ideal protocol (black solid line) is plotted against the diabatic error (red dotted line), obtained including a constant magnetic field in the Z direction of magnitude $\delta h_z=10^{-3} J$~\cite{Haffner08}, a $1\%$ error on the relative coupling magnitude $J_{12}/J_{23}$ and a NNN coupling $J_{13}=J_{12}/8$ .}
             \label{Fig1}
\end{figure}

Here we are interested in the pure Majorana regime in which $J\gg |h_z|$. The aim is to achieve this regime in the always-on interaction, such that the degenerate ground state of Hamiltonian~(\ref{KitaevSpinH}) encodes the topologically protected subspace. This degenerate ground state is composed of two Greenberger-Horne-Zeilinger (GHZ)-like states in the $X$ basis, which are highly entangled, $|\Psi_0\rangle=(|\leftarrow\leftarrow...\leftarrow\rangle-|\rightarrow\rightarrow...\rightarrow\rangle)/\sqrt{2}$, and $|\Psi_1\rangle=(|\leftarrow\leftarrow...\leftarrow\rangle+|\rightarrow\rightarrow...\rightarrow\rangle)/\sqrt{2}$. These states have different parity $P=-\prod_i\sigma^z_i$, $P|\Psi_{0,1}\rangle=\pm|\Psi_{0,1}\rangle$, which is a conserved quantity with respect to Hamiltonian~(\ref{KitaevSpinH}). On the other hand, they can also be easily achieved using an adiabatic evolution~\cite{Farhi00} from the opposite regime in which $J\ll |h_z|$. The ground state of the Hamiltonian $-h_z\sum_j\sigma_j^z$ for $h_z<0$, $|\Psi_0^{\rm init}\rangle=|\!\!\downarrow\downarrow...\downarrow\rangle$ is mapped adiabatically onto the ground state $|\Psi_0\rangle$ of the Hamiltonian $-J\sum_{j=1}^{N-1}\sigma_j^x\sigma_{j+1}^x$ for odd $N$, and to $|\Psi_1\rangle$ for even $N$. In turn, one specific linear combination of the first excited states of  $-h_z\sum_j\sigma_j^z$, $|\Psi_1^{\rm init}\rangle=\sum_i c_i|\downarrow\downarrow...\uparrow_i...\downarrow\rangle$, is mapped onto the other ground state. For instance, for $N=3$, $c_{1}=c_{3}=1/2$, and $c_{2}=1/\sqrt{2}$. Notice, in addition, that the two states will not be mixed during the adiabatic evolution due to their different parity. Despite the closure of the energy gap between the ground and the first excited states, parity commutes with the Hamiltonian at all times during the protocol, such that the dynamics will not mix the two states. This is the reason for the small diabatic error, plotted as a function of time in Fig.~\ref{Fig1}b. Moreover, for three qubits, and for adiabatic enough protocol, the first excited state is not mixed with other single-excitation states, giving similar fidelities than for the ground state, with an adiabatic protocol time similar to the one shown in Fig.~\ref{Fig1}b.

 In order to create state $|\Psi_0^{\rm init}\rangle$ one may just apply standard optical pumping. In order to create $|\Psi_1^{\rm init}\rangle$ one may apply an inhomogeneous Tavis-Cummings Hamiltonian~\cite{TavisCummings} with appropriate couplings, engineered with an inhomogeneous red-sideband interaction upon the ions, using the center-of-mass mode (e.g., with displaced equilibrium positions in the trap by a tailored potential~\cite{SchmiedLeibfried}). An important point here is that the Hamiltonian in the topological regime has to be turned on all the time to guarantee the topological protection against local noise, as we will explain in the following.

\section{Quantum Interface}
In addition to state initialization, the adiabatic mapping can be also used to transfer an arbitrary qubit state that comes into the system, e.g., a photonic qubit, to the MF qubit. One can begin with a photonic incoming state, $\alpha|0\rangle+\beta|1\rangle$, where $|0\rangle$, $|1\rangle$ are Fock states: by using a quantum interface with a cavity, state $|0\rangle$ can be mapped to $|\Psi_0^{\rm init}\rangle$ and $|1\rangle$ to $|\Psi_1^{\rm init}\rangle$, by a collective excitation and an intermediate phonon state~\cite{Lamata11b}. We point out that in order to get high fidelities in the coupling to the incoming photon, chains of around 20 ions at least should be employed~\cite{Lamata11b}, though for the MF qubit itself, as we will show, 3 ions are enough. In this respect, some experiments have already been realized~\cite{Cetina13,Stute}. Subsequently, the adiabatic transfer will produce the same qubit superposition $\alpha|\Psi_0\rangle+\beta| \Psi_1\rangle$ in the MF qubit. An extension could be considered with $K$ copies of this system, with a highly-entangled incoming photonic state,  $\sum c_{i_1...i_K}|i_1...i_K\rangle$. Here each photonic qubit is mapped similarly to the corresponding MF qubit, giving  $\sum c_{i_1...i_K}|\Psi_{i_1}...\Psi_{i_K}\rangle$, i.e., an arbitrarily entangled state of nonlocal Majorana-fermion qubits. This could be used as an efficient quantum memory. In Fig. \ref{Fig1}a we show a scheme of the proposed setup.

\section{Local operations and measurement}
One can efficiently implement local operations upon the MF qubits. In general, these local operations upon the nonlocal MF qubits translate into nonlocal operations upon the physical qubits, i.e., the trapped ions. The complete set of local operations, i.e. Pauli matrices, upon the nonlocal MF qubit, $\sigma^x_{\rm MFQ}$,  $\sigma^y_{\rm MFQ}$, and $\sigma^z_{\rm MFQ}$, can be related to the single-ion Pauli operators $\{\sigma^{x,y,z}_i\}_{i=1,..,N}$ through the expressions
\begin{eqnarray}
\label{rotation}
&&\sigma_{\rm MFQ}^x\equiv a^\dag_M+a_M=I\otimes I\otimes...\otimes I\otimes \sigma^x_1,\nonumber\\  
&&\sigma_{\rm MFQ}^y\equiv -i(a^\dag_M-a_M)=-\sigma^y_N\otimes \sigma^z_{N-1}\otimes...\otimes  \sigma^z_1,\label{LocalGatesMajorana}\\ \nonumber 
&&\sigma_{\rm MFQ}^z\equiv a^\dag_Ma_M-a_Ma^\dag_M = \sigma^y_N\otimes \sigma^z_{N-1}\otimes...\otimes \sigma^z_2\otimes \sigma^y_1.
\end{eqnarray}
As can be appreciated, $\sigma_{\rm MFQ}^{y,z}$ are highly nonlocal operations upon the ions. Nervertheless, they can be implemented efficiently with a reduced number of lasers as recently shown both theoretically~\cite{CasanovaFermions,Mueller11}, and experimentally~\cite{Barreiro11}, using sequences of  M\o lmer-S\o rensen and local gates. Thus, with the proposed setup, we have a fully controllable Majorana-fermion qubit.

The topological qubit readout can be also implemented.  A projective measurement upon the basis $\{|\Psi_0\rangle,|\Psi_1\rangle\}$, which is equivalent to detecting the parity operator $P=-\prod_i\sigma
^z_i$, amounts to one local measurement of $\sigma^z$ operator per ion. This is the easiest detection performed in trapped ions and can be done, with standard resonance fluorescence, with fidelities larger than 0.99~\cite{Leibfried03}. Combined with the local operations exposed above, this allows one for the full tomographic reconstruction of the MF qubit.

\section{Errors and decoherence protection}
The proposed implementation, as has been shown above, contains a degenerate ground state, $\{|\Psi_0\rangle$, $|\Psi_1\rangle\}$, which is protected by a gap from higher-level excitations.  
A consequence of this is that local operations $\sigma_i^{y,z}$, which couple  $\{|\Psi_0\rangle$, $|\Psi_1\rangle\}$ with higher energy states, are topologically supressed. Thus, magnetic field fluctuations along these directions will not have an effect upon the system. On the other hand, local $\sigma_i^x$ operations realize swap gates between the $|\Psi_0\rangle$ and $|\Psi_1\rangle$ states~\cite{KitaevLaumann}.  In a trapped-ion setup, random ambient magnetic fields along X direction rotate fast in the interaction picture with respect to the trapped-ion qubit energy in which Eq. (3) is computed~\cite{Kim09} , such that their contribution to decoherence will be negligible. This is because this spurious interaction is far off-resonant and will not induce transitions. 
For example, using a quadrupole transition between the two levels $|D_{5/2,3/2}\rangle$ and $|S_{1/2,-1/2}\rangle$ of $^{40}$Ca$^+$, with an energy separation of about $\omega_0=411$THz, an X error of magnitude $J\cdot10^{-3}$, e.g. around $6$Hz, will rotate in the interaction picture, taking the form $J\cdot10^{-3}(\sigma^+e^{i\omega_0t}+\sigma^-e^{-i\omega_0t})$, resulting in an effective unintended excitation probability of about $10^{-28}$.
In general this will also happen with the $Y$-direction ambient magnetic fields, such that these will be doubly protected: by the topological Hamiltonian and by the single-ion qubit energy transition. The local-noise protection scheme is summarized in Table \ref{ProtectionTable}. Conversely, local rotations in Eq. (\ref{rotation}) are realized in interaction picture with respect to the ion energy and commute with the Hamiltonian in the topological regime, also in the presence of non-nearest-neighbour (NNN) couplings, i.e. they can be realized efficiently. 

\begin{figure}[t] 
\includegraphics[width=0.9\linewidth]{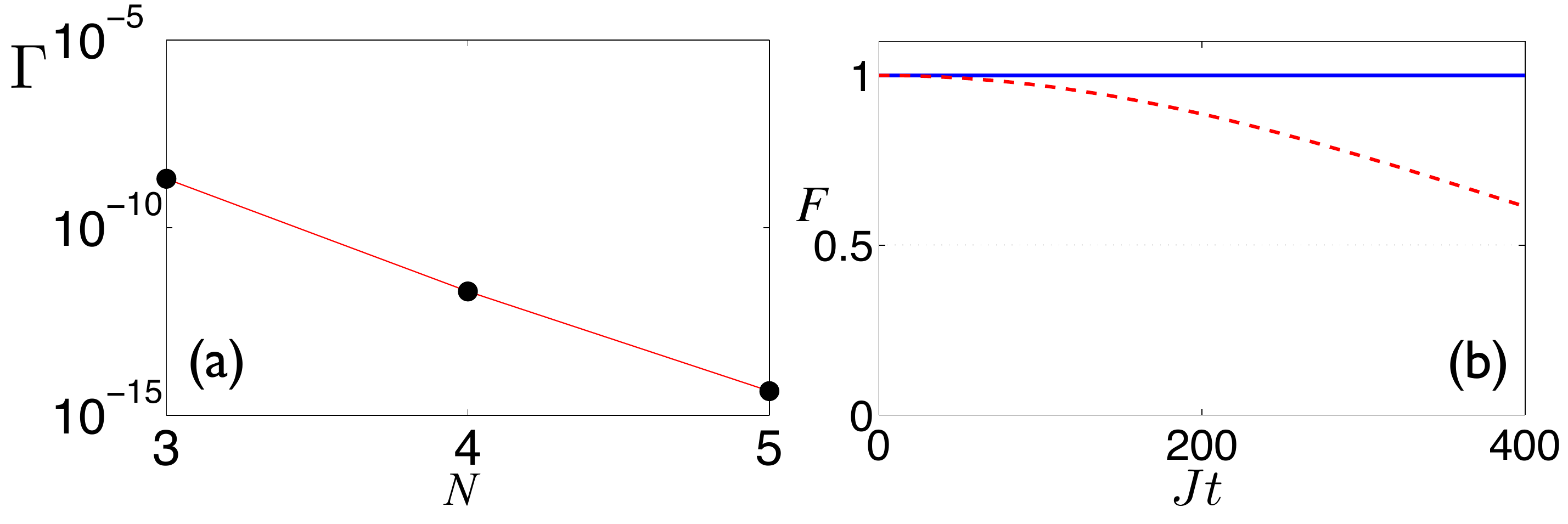}\centering
\caption{(a) Energy splitting $\Gamma$  between $|\Psi_0\rangle$ and $|\Psi_1\rangle$ as a function of  number of sites $N$, in units of $J=1$, computed numerically for $\delta h_z=10^{-3}J$ and $1\%$ error on the relative magnitude $J_{12}/J_{23}$. (b) Survival probability $F=|\langle\Psi_0|\psi(t)\rangle|^2$ for the evolved state $|\psi(t)\rangle$ from $|\Psi_0\rangle$ under the dynamics of Hamiltonian (\ref{KitaevSpinH}) (with $h_z=0$) (solid) or without it, i.e. free evolution of the state (dashed), plus constant local operators of modulus $10^{-3}J$, proportional to $\sigma_i^z$, for $N=3$. The topological Hamiltonian provides protection against the $\sigma_i^z$ noise.} 
             \label{Fig2}
\end{figure}

\begin{table}
\caption{\label{ProtectionTable}State protection against local noise sources in X,Y,Z directions.}
\footnotesize\rm
\begin{tabular*}{\textwidth}{@{}l*{15}{@{\extracolsep{0pt plus12pt}}l}}
\br
&X&Y&Z\\
\mr
Topological gap&-&\checkmark&\checkmark\\
Ion internal transition&\checkmark&\checkmark&-\\
\br
\end{tabular*}
\end{table}
A qubit encoded in the Kitaev wire model~\cite{KitaevWire01} is usually considered robust in the large $N$ limit, for the parameter range $|\mu|\ll 2w$. On the other hand, for finite $N$, the degree of imperfection in the protocol depends on how much the system deviates from the parameter regime $w=|\Delta|$, $\mu=0$. Indeed, the appearing energy splitting $\Gamma$ between $|\Psi_0\rangle$ and $|\Psi_1\rangle$, which breaks the ground-state degeneracy and qubit coherence~\cite{KitaevWire01}, is of the order of $\Gamma\propto \exp(-N/n_0)$, where $n_0^{-1}=\min\{|\ln|x_+||,|\ln|x_-||\}$, and
$x_{\pm}=(-\mu\pm\sqrt{\mu^2-4w^2+4|\Delta|^2})/[2(w+|\Delta|)]$.
We consider  $J/2\pi\sim 6$ kHz for the 3 ion case. For realistic imperfections in the Rabi frequencies of the lasers of 1$\%$, that induce the same order of imperfection in $w=|\Delta|=J$, and magnetic field fluctuations in $\delta h_z=10^{-3}J$~\cite{Haffner08}, we have $n_0\simeq 0.14$. The splitting computed numerically  as a function of the number of sites is plotted in Fig. (\ref{Fig2}). This makes the  implementation of the wire model and Majorana fermions in trapped ions a real possibility with current technology. We show below that considering a linear chain of just $N=3$ ions, one may encode a topologically-protected qubit subspace with very low decoherence. 

In order to test the feasibility of the implementation with realistic trapped-ion systems, we have realized numerical simulations with different kinds of imperfections. With respect to the adiabatic protocol with $H_s$ in Eq.~(\ref{KitaevSpinH}), we have computed the fidelity loss  $1-F(t)=1-|\langle\Psi_0|\psi(t)\rangle|^2$ for the evolved state $|\psi(t)\rangle$ from $|\Psi_0^{\rm init}\rangle$, making the evolution $|h_z|/J\gg 1\rightarrow |h_z|/J\ll1$, for $N=3$, see Fig. \ref{Fig1}b. We consider the ideal case with no ambient magnetic field and coupling errors, and the case with a $\delta h_z=10^{-3}J$ constant magnetic field in the Z direction, a NNN coupling $J_{13}=J_{12}/8$ and a $1\%$ error on the relative magnitude $J_{12}/J_{23}$. The diabatic error in both cases is of the order $10^{-3}$. We plot in Fig. \ref{Fig2} (a)  the energy splitting $\Gamma$ between $|\Psi_0\rangle$ and $|\Psi_1\rangle$  as a function of $N$, in units of $J=1$, computed numerically  for $\delta h_z=10^{-3}J$ and a $1\%$ error on the relative magnitude $J_{12}/J_{23}$. The scaling is exponential, as expected. For $N=3$,  we have $\Gamma\simeq 2\times 10^{-9}J$, i.e., a negligible dephasing error. Thus, the qubit encoding will in principle work already for 3 ions, which is feasible with current technology.

We remark that the presence of a NNN coupling does not affect the protocol for a three-ion setup. A  spurious coupling $J_{13}$ between the first and the third ions does not modify the protocol in terms of ground subspace (the ground subspace  ${\rm span}\{|\Psi_0\rangle$, $|\Psi_1\rangle\}$ stays the same), in terms of splitting $\Gamma$ (in presence of a $J_{13}$ the energies of  $|\Psi_0\rangle$ and $|\Psi_1\rangle$ only show a coherent down shift) and in terms of local operations upon the qubit. Thus even by addressing the three ions with one bichromatic laser is sufficient to implement the model~\cite{Kim09}. Nevertheless, the Hamiltonian free of NNN coupling for three sites can be obtained by using different detuned lasers, as shown in ~\cite{Mezzacapo12}.

To show the subspace topological protection of the degenerate ground state $\{|\Psi_0\rangle$, $|\Psi_1\rangle\}$ with respect to coupling to outer states, we plot in Fig. \ref{Fig2} (b) the survival probability $|\langle\Psi_0|\psi(t)\rangle|^2$ for the evolved state $|\psi(t)\rangle$ from $|\Psi_0\rangle$ under the dynamics of Hamiltonian (\ref{KitaevSpinH}) (solid) or without it (dashed) with a constant magnetic field $\delta h_z\sim 10^{-3}J$ along the Z axis for $N=3$. This is the kind of local operators that couple  ${\rm span}\{|\Psi_0\rangle$, $|\Psi_1\rangle\}$ with outer states. It is clearly appreciated that the survival probability inside the subspace is significantly increased upon evolution with (\ref{KitaevSpinH}). We remark that this is the main fundamental decoherence source in most experimental trapped ion setups~\cite{Haffner08}, with coherence lifetimes of about 30 ms. We have that states $|\Psi_0\rangle$, $|\Psi_1\rangle$ will couple to each other or to outer states only through higher-order processes in perturbation theory. The largest contribution to decoherence is at most of order $\sim\delta h_z^2/\Delta_g$. Here $\delta h_z$ is the average local magnetic field perturbation, that we take, as in the numerical simulation, to be $10^{-3}J$, and $\Delta_g=2J$ is the gap between the topological ground states $|\Psi_0\rangle,|\Psi_1\rangle$ and the excited states. For this we have assumed that these spurious magnetic fields change in time much more slowly than the frequency of the gap. Accordingly, the effective Rabi frequency of the error is about $5\times 10^{-4}\delta h_z$, i.e., several orders of magnitude reduced with respect to the case without topological protection [without evolution by (\ref{KitaevSpinH})], that is of order $\delta h_z$. We point out that, in the long time limit, other sources of error will dominate on the decoherence of the system. The choice for the optimal size of the ion array, in which one encodes the topological qubit, will depend on the type of errors and parameter control of the particular experimental setup. Indeed, while for shorter chains one has less errors due to spontaneous emission, for longer chains the realistic ground state energy splitting between the qubit basis states will be smaller. For a three-ion array, considering that motional heating rates in trapped ions can be of a few phonons per second and spontaneous emission lifetimes of more than $1$s~\cite{Haffner08}, with this proposal one may improve coherence lifetimes by more than one order of magnitude with current technology.

\section{Conclusions}
We have shown that trapped-ion chains can host a topologically protected qubit subspace by means of MFs. We predict that MF qubits, encoded in a chain of three ions, can outperform the usual ionic qubit coherence time by more than one order of magnitude, yielding an efficient quantum memory. Local rotations upon the qubit can be performed by means of global and local laser-ion interactions. Moreover, a quantum interface with photonic states can be realized, allowing for the realization of two-qubit gates among the MF qubits and quantum communication.

\subsection*{Acknowledgments}

The authors acknowledge funding from the Basque Government grant IT472-10, an EC Marie
Curie IEF grant, the Spanish MINECO grant FIS2012-36673-C03-02, UPV/EHU UFI 11/55,
and the EU projects SOLID, PROMISCE, CCQED and SCALEQIT.

\end{document}